\documentclass[aps,pra,floatfix,amsmath,amssymb,eqsecnum,nofootinbib,twocolumn]{revtex4}
\usepackage{graphicx}
\usepackage{dcolumn}
\usepackage{bbm}
\usepackage{epstopdf}
\usepackage[bookmarks=true,colorlinks,linkcolor=blue,urlcolor=blue,citecolor=blue]{hyperref}
\usepackage{color}

\usepackage{setspace} 
\newcommand{\beq}{\begin{equation}}
\newcommand{\eeq}{\end{equation}}
\newcommand{\beqn}{\begin{eqnarray}}
\newcommand{\eeqn}{\end{eqnarray}}

\usepackage{amsmath}
\usepackage{amssymb}

\def \q{{\mathbf{q}}}
\def \R{{\mathbf{R}}}

\def \k{{\mathbf{k}}}

\def \q{{\mathbf{q}}}
\def \p{{\mathbf{p}}}

\def \R{{\mathbf{R}}}

\def \S{{\mathbf{S}}}

\def \tn{\textnormal}

\begin{document}

\title{Topological excitations and the dynamic structure factor \\ of spin liquids on the kagome lattice}

\author{Matthias Punk}
\affiliation{Department of Physics, Harvard University, Cambridge MA
02138}

\author{Debanjan Chowdhury}
\affiliation{Department of Physics, Harvard University, Cambridge MA
02138}

\author{Subir Sachdev}
\affiliation{Department of Physics, Harvard University, Cambridge MA
02138}

\date{\today }

\begin{abstract}

Recent neutron scattering experiments on the spin-1/2 kagome lattice antiferromagnet ZnCu$_3$(OH)$_6$Cl$_2$ (Herbertsmithite) 
provide the first evidence of fractionalized excitations in a quantum spin liquid state in two spatial dimensions. 
In contrast to existing theoretical models
of spin liquids, the measured dynamic structure factor reveals an excitation continuum which is remarkably flat as a function of frequency and has almost no momentum dependence along several high-symmetry directions.
Here we show that many experimentally observed features can be explained by the presence of topological {\em vison\/} excitations in a $Z_2$ spin liquid. These visons form flat bands on the kagome lattice, and thus act as a momentum sink for spin-carrying excitations which are probed by neutron scattering. We compute the dynamic structure factor for two different $Z_2$ spin liquids and find that one of them describes Herbertsmithite well above a very low energy cutoff.

\end{abstract}

\maketitle

\section{Introduction}

The search for quantum spin liquid states in frustrated magnets remains a very active area of research in condensed matter 
physics \cite{Balents}. One reason is that these novel quantum phases provide an ideal basis to study exotic states of matter which support fractionalized excitations.
Herbertsmithite, a layered spin-1/2 kagome lattice antiferromagnet \cite{Shores}, is one of the strongest contenders for an experimental realization of a spin liquid state \cite{deVries}. Indeed, no sign of magnetic ordering is observed down to temperatures around 50mK, while the natural energy scale set by the magnetic exchange coupling $J \sim 200$K is four orders of magnitude larger \cite{Helton}.  

Neutron scattering experiments \cite{Han} on single crystals of this material are consistent with a continuum of fractionalized spinon excitations as expected in a quantum spin liquid state. Remarkably, the measured dynamic structure factor shows hardly any momentum dependence along several high-symmetry directions.
By contrast, all mean-field models with spinons \cite{Subir1992,Wang,Ran,Poilblanc,Motrunich,Oleg}, whether gapless or gapped, 
give rise to sharp dispersing features in the dynamic structure factor \cite{Messio, Dodds} 
which have not been observed. In particular, theory predicts a vanishing structure factor below the onset of the two spinon continuum, which is at a finite energy even for gapless spin liquids, apart from the small set of crystal momenta where the spinon gap closes. This is in stark contrast to experiments, where the measured structure factor is finite and almost constant as a function of frequency down to energies on the order of $\sim J/10$ \cite{Han}.

Here we propose an explanation for the lack of a momentum-dependent spinon continuum threshold via 
the interaction of spinons with another set of excitations which form a (nearly) {\em flat\/} band. Such localized excitations act as a momentum sink for the spinons, thereby flattening the dynamic structure factor.
So far, the only theoretical model for a spin liquid state on the kagome lattice which naturally gives rise to a flat excitation band 
at low energies are the $Z_2$ spin liquids \cite{Subir1992,Wang,Ran}. 
Besides spinons, these states exhibit gapped vortex excitations \cite{rc,kivelson} of an emergent $Z_2$ gauge field \cite{rstl,wen1}, 
so called visons \cite{sf}, which indeed have a lowest energy band which is nearly flat \cite{Nikolic,Huh}. Visons carry neither charge nor spin and thus do not couple directly to neutrons. They interact with spinons, however, and we show that this coupling is responsible for flattening the dynamic structure factor and removing the sharp onset at the two-spinon continuum, in accordance with experimental results.
Note that the vison gap has to be small for this mechanism to work. This assumption is justified by numerical density matrix renormalization group calculations \cite{White,Balents2,Schollwoeck}, which indicate that a $Z_2$ spin liquid ground-state on the kagome lattice is proximate to a valence bond solid (VBS) transition. The vison gap, which closes at the transition to a VBS phase, is thus expected to be small.

\begin{figure*}
\begin{center}
\includegraphics[width=0.9 \textwidth]{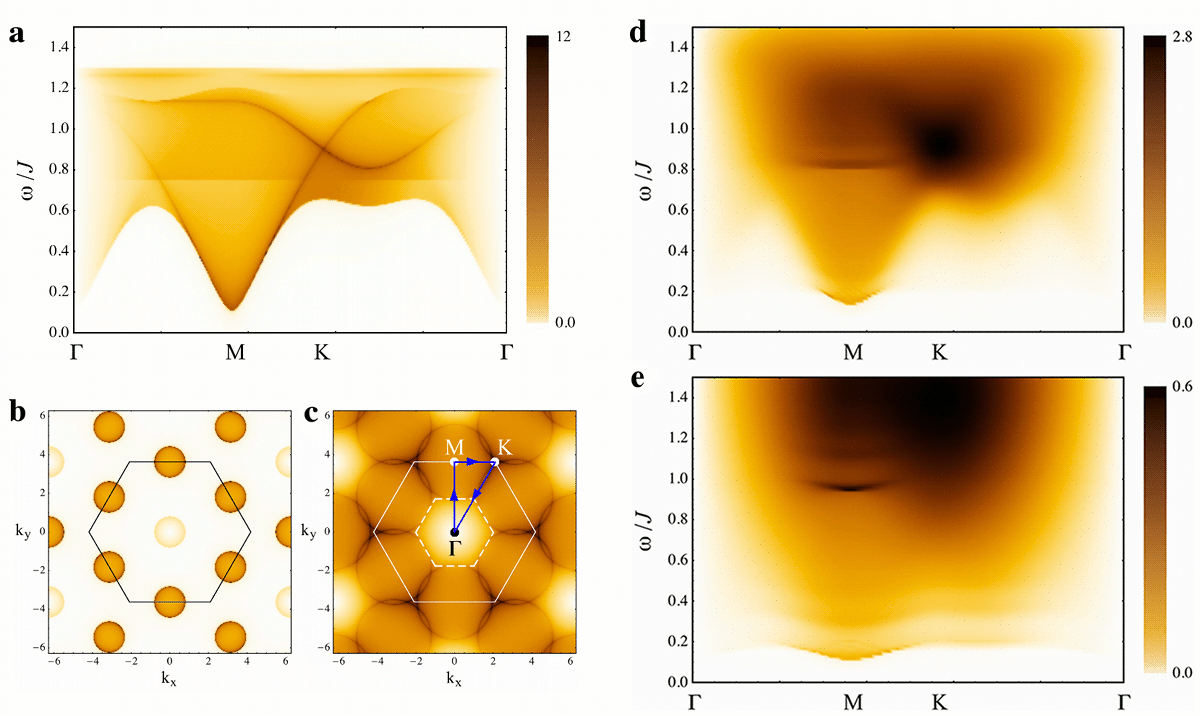}
\end{center}
\caption{Density plots of the the dynamic spin-structure factor $S(\k,\omega)$ for the $Q_1=Q_2$ spin liquid state. (a), (d) and (e) show $S(\k,\omega)$ for different spinon-vison interaction strengths as a function of frequency and momentum along high symmetry directions between the $\Gamma$, $M$ and $K$ points of the extended Brillouin zone, indicated by the blue arrows in (c). Panel (a): non-interacting spinons. Note that in the $Q_1=Q_2$ state two of the three spinon bands are degenerate, whereas the third, highest energy spinon band is flat. This flat spinon band gives rise to the horizontal feature at $\omega \simeq 0.75 J$ in (a).  (d): spinon-vison interaction $g_0=0.2$, (e): spinon-vison interaction $g_0=0.6$. Panels (b) and (c) show $S(\k,\omega)$ for non-interacting spinons at fixed frequency $\omega/J=0.4$ (b) and $\omega/J = 0.85$ (c). The elementary Brillouin zone of the kagome lattice is indicated by a dashed hexagon in (c).
Note the sharp onset of the two-spinon continuum for non-interacting spinons in (a) and (b), which is washed out when interactions with visons are accounted for. 
All data in this figure was calculated for $|Q_1|=0.4$ and the spinon gap was fixed at $\Delta_s \simeq 0.05 J $. The vison gap is set to $\Delta_v = 0.025 J$ in (d) and (e). }
\label{fig1}
\end{figure*}

\section{Model}

Our aim is to compute the dynamic structure factor for two $Z_2$ spin liquids which have been discussed in detail in Ref.~\onlinecite{Subir1992}. 
We start from the standard bosonic spin liquid mean-field theory of the spin-1/2 antiferromagnetic Heisenberg model on the kagome lattice. Using a Schwinger-boson representation of the spin-1/2 operators $\mathbf{S}_i = b^\dagger_{i \alpha} \boldsymbol{\sigma}_{\alpha \beta} b^{\ }_{i \beta} / 2$ and performing a mean-field decoupling in the spin-singlet channel, the Heisenberg Hamiltonian can be written as
\begin{equation}
H_b = - J \sum_{\langle i,j \rangle} Q^*_{ij} \, \varepsilon_{\alpha\beta} \, b_{i\alpha} b_{j\beta} + \text{h.c.}  + \lambda \sum_i b^\dagger_{i\alpha} b^{\ }_{i\alpha} \ ,
\end{equation}
with $Q^*_{ij} = \langle \varepsilon_{\alpha\beta} \, b^\dagger_{i\alpha} b^\dagger_{j\beta} \rangle /2$ and $\lambda$ denotes the Lagrange multiplier which fixes the constraint of one Schwinger boson per lattice site. $\epsilon_{\alpha\beta}$ is the fully antisymmetric tensor of $SU(2)$.
As mentioned in the introduction, we want to study the effect of vison excitations on the spinons, thus we have to include phase fluctuations of the mean field variables $Q_{ij}$ in our theory. The $Z_2$ spin liquid corresponds to the Higgs phase of the resulting emergent gauge theory, where the phase fluctuations are described by an Ising bond variable $\sigma^z_{ij}$.
The Hamiltonian describing bosonic spinons and their coupling to the Ising gauge field takes the form
\begin{eqnarray}
H &=& - J \sum_{\langle i,j \rangle} \sigma^z_{ij} \big( Q^*_{ij} \, \varepsilon_{\alpha\beta} \, b_{i\alpha} b_{j\beta} + \text{h.c.} \big) + \lambda \sum_i b^\dagger_{i\alpha} b^{\ }_{i\alpha}   \notag \\
&& + K \sum_\text{plaq.} \prod_\text{plaq.} \sigma^z_{ij} - h \sum_{\langle i,j\rangle} \sigma^x_{ij} \ ,
\label{Heff}
\end{eqnarray}
where the terms in the second line are responsible for the dynamics of the gauge field $\sigma^z_{ij}$.
Vison excitations are vortices of this emergent $Z_2$ gauge field, i.e.~excitations where the product $ \prod  \sigma^z_{ij} $  on a plaquette changes sign. In order to study the interaction between spinons and visons it is more convenient to switch to a dual description of the $Z_2$ gauge field in terms of its vortex excitations \cite{Wegner}.
In this dual representation the pure gauge field terms in the second line of Eq.~\eqref{Heff} take the form of a fully-frustrated Ising model on the dice lattice, which has been studied in detail in Refs.~\onlinecite{Nikolic} and \onlinecite{Huh}. Within a soft-spin formulation restricted to nearest neighbor hopping, this model gives rise to three flat vison bands separated by an energy on the order of the exchange coupling $J$. Since only the gap to the lowest vison band is small, we neglect effects of the other two bands in the following.

The coupling between spinons and visons is a long-range statistical interaction (a spinon picks up a Berry's phase of $\pi$ when encircling a vison \cite{Huh}), which cannot be expressed in the form of a simple local Hamiltonian in the vortex representation.
However, the fact that visons on the dice lattice are non-dispersing comes to the rescue here. Since these excitations are localized and can only be created in pairs, the long-range statistical interaction is effectively cancelled. Indeed, if a spinon is carried around a pair of visons, it does not pick up a Berry's phase. For this reason it is reasonable to replace the long range statistical interaction by a local energy-energy coupling. Accordingly we choose the simplest, gauge-invariant Hamiltonian of bosonic spinons on the kagome lattice coupled to a single, non-dispersing vison mode on the dual Dice lattice 
\begin{eqnarray}
H &=& H_b + \sum_i \Delta_v \phi_i \phi_{i} \notag \\
&& + \, g_0 \Delta_v \sum_{ \substack{i  \in \text{Dice}_3 \\ \ell, m \in \bigtriangledown_i} } \phi_i \phi_i  \left( \varepsilon_{\alpha \beta} Q^*_{\ell m} b_{\ell \alpha} b_{m \beta}  + \text{h.c.} \right) \, . \ \ \ \
\label{H_SV}
\end{eqnarray}
Here, the real field $\phi_i$ describes visons living on the dice lattice sites $i$ and $\Delta_v$ is the vison gap. The sum in the interaction term runs only over the three-coordinated Dice lattice sites $i$ and couples the spinon bond energy on the triangular kagome plaquettes to the local vison gap at the plaquette center. Further terms, where spinons on the hexagonal kagome plaquettes interact with visons at the center of the hexagons are allowed, but neglected for simplicity.

A more detailed discussion of this interaction term can be found in the supplementary material. 
We are going to compute the dynamic structure factor $S(\k,\omega)$ using the model \eqref{H_SV} for a particular $Z_2$ spin liquid state which has been identified in Ref.~\onlinecite{Subir1992}. For the nearest neighbor kagome antiferromagnet there are basically two independent bond expectation values $Q_{ij}\in \{Q_1,Q_2\}$ and the two distinct, locally stable mean-field solutions have $Q_1=Q_2$ or $Q_1=-Q_2$. The $Q_1=Q_2$ state has flux $\pi$ in the elementary hexagons, whereas the $Q_1=-Q_2$ state is a zero-flux state. During the remainder of this article we focus only on the $Q_1=Q_2$ state, since it gives rise to a little peak in $S(\k,\omega)$ at small frequencies at the $M$ point of the extended Brillouin zone, in accordance with experimental results. The presence of this peak is related to a minimum in the two-spinon continuum at the $M$ point (see Fig.~\ref{fig1}). By contrast, the two-spinon continuum for the $Q_1=-Q_2$ state has a minimum at the $K$ point (see supplementary information). Two other bosonic $Z_2$ states have been identified on the kagome lattice\cite{Wang}, but we refrain from computing the structure factor for these states, because both have a doubled unit-cell which complicates the calculations considerably.

Note that we do not determine the parameters $|Q_1|$ and $\lambda$ variationally. Instead, we use them to fix the spinon gap as well as the spinon bandwidth. $|Q_1|$ is restricted to values between $0$ and $1/\sqrt{2}$ and quantifies antiferromagnetic correlations of nearest neighbor spins ($|Q_1|=1/\sqrt{2}$ if nearest neighbor spins form a singlet). All data shown in this paper was computed for $|Q_1|=0.4$, and $\lambda$ has been adjusted such that the spinon gap takes the value $\Delta_s/J \simeq 0.05$. As mentioned in the introduction, we assume that the vison gap $\Delta_v$ is small due to evidence of proximity to a VBS state, and we chose $\Delta_v/J = 0.025$ for all data shown in this article, i.e.~the vison gap is roughly half the spinon gap.

\section{Dynamic structure factor}

Neutron scattering experiments measure the dynamic structure factor
\begin{equation}
S(\k,\omega) = \frac{1}{N} \sum_{i,j} e^{i \k \cdot (\R_i-\R_j)} \int dt \, e^{-i \omega t} \, \langle \S_i(t) \cdot \S_j(0) \rangle \ ,
\end{equation}
which we are going to compute for the model presented in Eq.~\eqref{H_SV}. Note that $S(\k,\omega)$ is periodic in the extended Brillouin zone depicted in Fig.~\ref{fig1} (c).

\begin{figure}
\begin{center}
\includegraphics[width=0.85 \columnwidth]{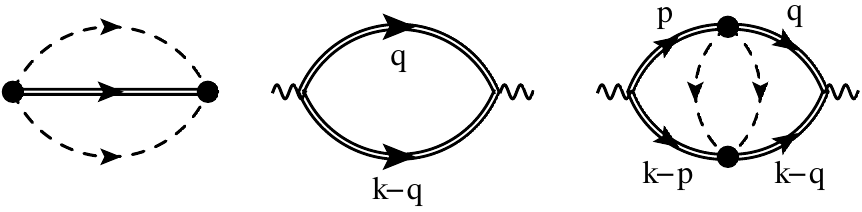}
\end{center}
\caption{Spinon self energy (left), one-loop contribution to the spin susceptibility (middle)  and corresponding lowest order vertex correction (right). Double lines are dressed spinon propagators and dashed lines are bare vison propagators.}
\label{figdiags}
\end{figure}

\begin{figure*}
\begin{center}
\includegraphics[width= \textwidth]{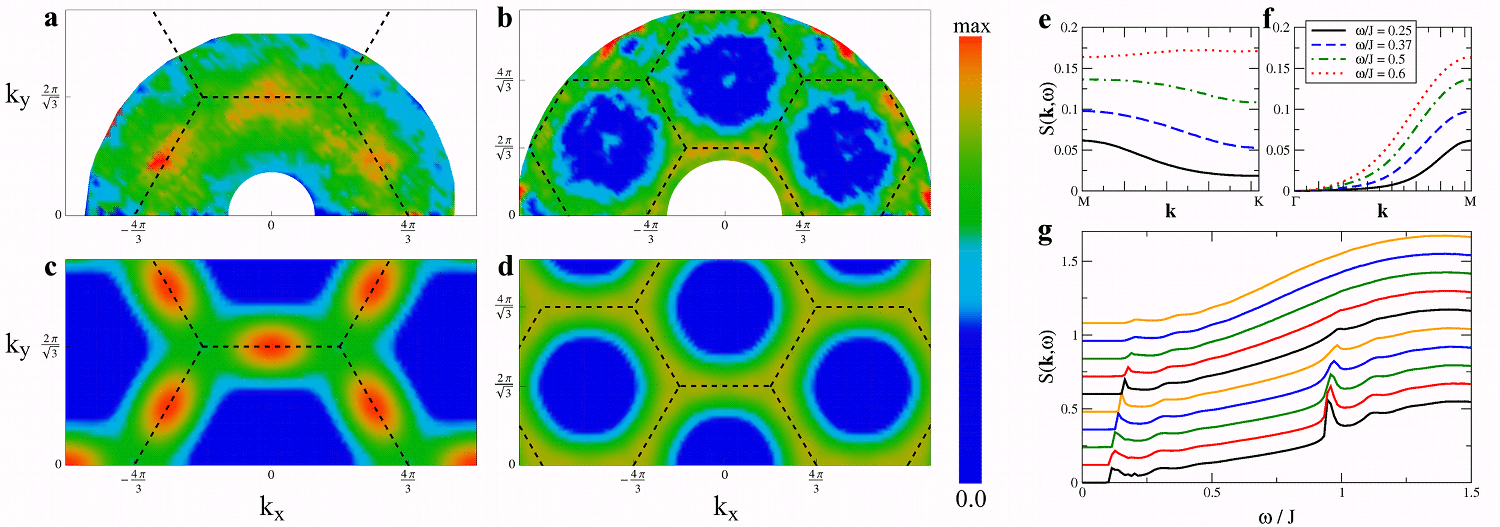}
\end{center}
\caption{Qualitative comparison between experimental measurements \cite{Han} and our theoretical results for the dynamic structure factor $S(\k,\omega)$. Experimental data at fixed frequency are shown for (a) $\omega = 0.75$meV and (b) $\omega= 6$meV. Theoretical results for the $Q_1$=$Q_2$ spin liquid at fixed frequency are plotted for (c) $\omega= 0.37J$ and (d) $\omega=0.6J$. The extended Brillouin zone is indicated by the dashed hexagons. Note that the peak at the $M$ point at low frequencies, as well as the flatness of $S(\k,\omega)$ between the $M$ and $K$ points at higher frequencies is captured by our theory. Cuts of our theoretical results for $S(\k,\omega)$ along high symmetry directions at different frequencies are plotted in (e) between the $M$ and $K$ point, as well as in (f) between the $\Gamma$ and $K$ point, again showing the peak at the $M$ point at low frequencies.
Panel (g) shows details of the calculated structure factor as function of frequency for various momenta between the $M$ (bottom curve) and $K$ point (top curve). Note that all curves in (g) are shifted by $0.12 J$ with respect to each other for better visibility. All theoretical data shown was computed for the $Q_1=Q_2$ state with a spinon-vison interaction strength $g_0=0.6$ and other parameters as in Fig.~\ref{fig1}.}
\label{fig3}
\end{figure*}

After expressing $\S_i \cdot \S_j$ in terms of Schwinger bosons and diagonalizing the free spinon Hamiltonian with a Bogoliubov transformation, the one loop expression for the dynamic spin-susceptibility can be derived straightforwardly and takes the form
\begin{eqnarray}
\chi(\k,i \omega_n) &=& \frac{3}{2} \sum_{\q,\Omega_n} G_\ell(\q,i \Omega_n) G_m(\k-\q, i\omega_n-i\Omega_n) \notag \\
&& \times \Big[ U_{j\ell}(\q) V_{jm}(\k-\q) + V_{j\ell}(\q) U_{jm}(\k-\q) \Big] \notag \\
&& \times \, U^*_{i\ell}(\q) V^*_{im}(\k-\q)  + \dots \ ,
\label{suscept}
\end{eqnarray}
where the dots represent similar terms which give a contribution at negative frequencies after analytic continuation and thus play no role for calculating $S(\k,\omega)$ at zero temperature. The summation over the sublattice indices $i,j,\ell, m \in \{1,2, 3\}$ is implicit here and the $3\times 3$ matrices $U_{ij}$ and $V_{ij}$ form the Bogoliubov rotation matrix
\begin{equation}
M = \begin{pmatrix}
U & -V^* \\
V & U^*
\end{pmatrix} \ ,
\end{equation} 
as defined in Ref.~\onlinecite{Subir1992}, which diagonalizes the mean-field spinon Hamiltonian.
$G_\ell(\q,i \Omega_n)$ denotes the dressed spinon Green's function with band-index $\ell$
\begin{equation}
G^{-1}_\ell(\q,i \Omega_n) = i \Omega_n -\epsilon_\ell(\q) -\Sigma_\ell(\q, i \Omega_n)
\end{equation}
The spinon self-energy (see Fig.~\ref{figdiags}), which we compute self-consistently, is determined by the equation 
\begin{equation}
\Sigma_\ell (\q, i \Omega_n) = \sum_{\p,m} \lambda^\dagger_{\ell m}(\p,\q) \lambda^{\ }_{m \ell}(\p,\q) G_m(i \Omega_n -2 \Delta_v,\p) \ .
\label{selfen}
\end{equation}
Here the $6\times6$ matrix $\lambda(\p,\q)$ denotes the bare spinon-vison interaction vertex, with $\p$ ($\q$) the momentum of the outgoing (incoming) spinon. 
Note that the six spinon bands come in three degenerate pairs due to the $SU(2)$ spin-symmetry. Furthermore, note that the flat vison band is not renormalized at arbitrary order in the spinon-vison coupling.

The dynamic structure factor  can be obtained from the susceptibility \eqref{suscept} via
\begin{equation}
S(\k,\omega) = \frac{ \text{Im} \, \chi(\k, i \omega_n \to \omega + i 0^+) }{1-e^{-\beta \omega}} \ .
\end{equation}
Results of this calculation at zero temperature are shown in Figs.~\ref{fig1} and \ref{fig3} for the $Q_1=Q_2$ state for different spinon-vison interaction strengths $g_0$. In the region around and in-between the high symmetry points $M$ and $K$ the lowest order vertex correction shown in Fig.~\ref{figdiags} gives only a relatively small contribution to $S(\k,\omega)$ and thus has been neglected in the data shown in these figures (see supplementary material for a discussion).

\section{Discussion}

Fig.~\ref{fig1} shows the two spinon contribution to the dynamic structure factor for the $Q_1=Q_2$ state (results for the $Q_1=-Q_2$ state can be found in the supplementary material). The onset of the two spinon continuum, which has a minimum at the $M$ point, is clearly visible in Fig.~\ref{fig1}(a) as the line of frequencies below which the dynamic structure factor vanishes. Moreover, several sharp peaks appear inside the spinon continuum. We note that such features in the two-spinon contribution to $S(\k,\omega)$ are generic and are present also for gapless Dirac spin liquids. Since none of these structures have been observed in experiment, it is clearly necessary to go beyond this level of approximation.

Figs.~\ref{fig1}(d) and (e) show the dynamic structure factor along the same high symmetry directions as in Fig.~\ref{fig1}(a), but now including the effect of spinon-induced vison pair production for two different interaction strengths $g_0$. The non-dispersing visons act as powerful momentum sink for the spinons and lead to a considerable shift of spectral weight below the two-spinon continuum. The computed structure factor is almost structureless and considerably flattened at intermediate energies. Our results for the $Q_1=Q_2$ state also capture the small low-frequency peak in $S(\k,\omega)$ at the $M$ point, which has been seen in experiment.
This peak is a remnant of a minimum in the threshold of the two-spinon continuum at the $M$ point, and we conjecture that it might be an indication that this particular $Z_2$ spin liquid state is realized in Herbertsmithite. In Fig.~\ref{fig3} we show plots of $S(\k,\omega)$ at constant energy, where this peak is clearly visible, and compare our results qualitatively to the experimental data.

In Figs.~\ref{fig1}(e) and \ref{fig3}(g) one can barely see small oscillations of $S(\k,\omega)$ at low frequencies. These oscillations originate from the self-consistent computation of the spinon self-energy $\Sigma(\k,\omega)$ and are related to resonances in the self-energy at energies corresponding to the creation of two, four, six, and higher even numbers of vison excitations. We emphasize here that a self-consistent computation of the spinon self-energy is necessary, because the real part of $\Sigma(\k,\omega)$ is large and broadens the spinon bands. A non-selfconsistent computation thus leads to sharp spinon excitations above the bare spinon band, which are unphysical as they would decay immediately via vison pair production. A different approximation, which circumvents this problem, would be to calculate $\Sigma(\k,\omega)$ non-selfconsistently and neglect the real part completely. This approximation violates sum-rules however, as the integrated spectral weight of the spinon is no longer unity (for a detailed discussion, see the supplementary material). 

Lastly, neutron scattering experiments explored energies up to $\omega\simeq0.65 J$ and concluded that the integrated weight accounts for roughly $20\%$ of the total moment sum rule \cite{Han}. Consequently it is reasonable to expect that the dynamic structure factor is finite up to energies of a few $J$. For the parameters chosen in our calculation (i.e. $Q_1=0.4$ and a spinon gap $\Delta_s \simeq 0.05$) the structure factor for non-interacting spinons has a sharp cutoff at an energy around $\omega \simeq 1.3 J$, corresponding to roughly twice the spinon bandwidth. If interactions with visons are included, this upper cutoff is shifted to considerably larger energies, however. For a spinon-vison coupling $g_0=0.6$, the structure factor has a smooth upper cutoff at an energy around $\omega \simeq 3J$. Such large bandwidths are hardly achievable in theories with non-interacting spinons. We note that similarly large bandwidths have been found in exact diagonalization studies \cite{Laeuchli}.

In conclusion, we've calculated the dynamic spin structure factor of a bosonic $Z_2$ spin liquid on the kagome lattice. Taking interactions with topological vison excitations into account, we showed that our results are in qualitative agreement with neutron scattering experiments above a low energy cutoff. Below this cutoff, there could be an instability to some other quantum state, and it is likely that impurity effects are also important in interpreting the
experiments.

We acknowledge illuminating discussions with M.~Babadi, S.~Gopalakrishnan, M.~Lawler,  J.~D.~Sau, and especially Y.~S.~Lee. 
Furthermore, we thank T.-H.~Han and Y.~S.~Lee for providing the neutron scattering data shown in Fig.~\ref{fig3}.
This research was supported by the US NSF under Grant DMR-1103860 and
by the John Templeton foundation. 
This research was also supported in part by Perimeter Institute for
Theoretical Physics; research at Perimeter Institute is supported by the
Government of Canada through Industry Canada and by the Province of
Ontario through the Ministry of Research and Innovation. The computations were performed in part on the Odyssey cluster supported by the FAS Science Division Research Computing Group at Harvard University.

\newpage

\begin{widetext}

\newpage

\section*{Supplementary material}

\subsection{Results for the $Q_1=-Q_2$ state.}

Here we briefly summarize our computations of the dynamic structure factor for the $Q_1=-Q_2$ state, shown in Fig.~\ref{figQ1mQ2}. In contrast to the $Q_1=Q_2$ state, the onset of the two-spinon continuum has a prominent minimum at the $K$ point in this case. Also note that $S(\k,\omega)$ for non-interacting spinons shown in Fig.~\ref{figQ1mQ2} reveals more structure than in the $Q_1=Q_2$ state, shown in Fig.~\ref{fig1}. This is because the $Q_1= -Q_2$ state has three distinct spinon bands, whereas the lower two spinon bands of the $Q_1=Q_2$ state are degenerate and the third band is flat. 
If interactions with visons are included, we expect a similar broadening of $S(\k,\omega)$ as discussed in the main text for the $Q_1=Q_2$ state, albeit without a small low-frequency peak at the $M$ point. Since our numerical algorithm for calculating the dynamical structure factor for interacting spinons is highly optimized for the $Q_1=Q_2$ state, we refrained from a computation including spinon-vison interactions for the $Q_1=-Q_2$ case.

\begin{figure}[b]
\begin{center}
\includegraphics[width=0.85 \columnwidth]{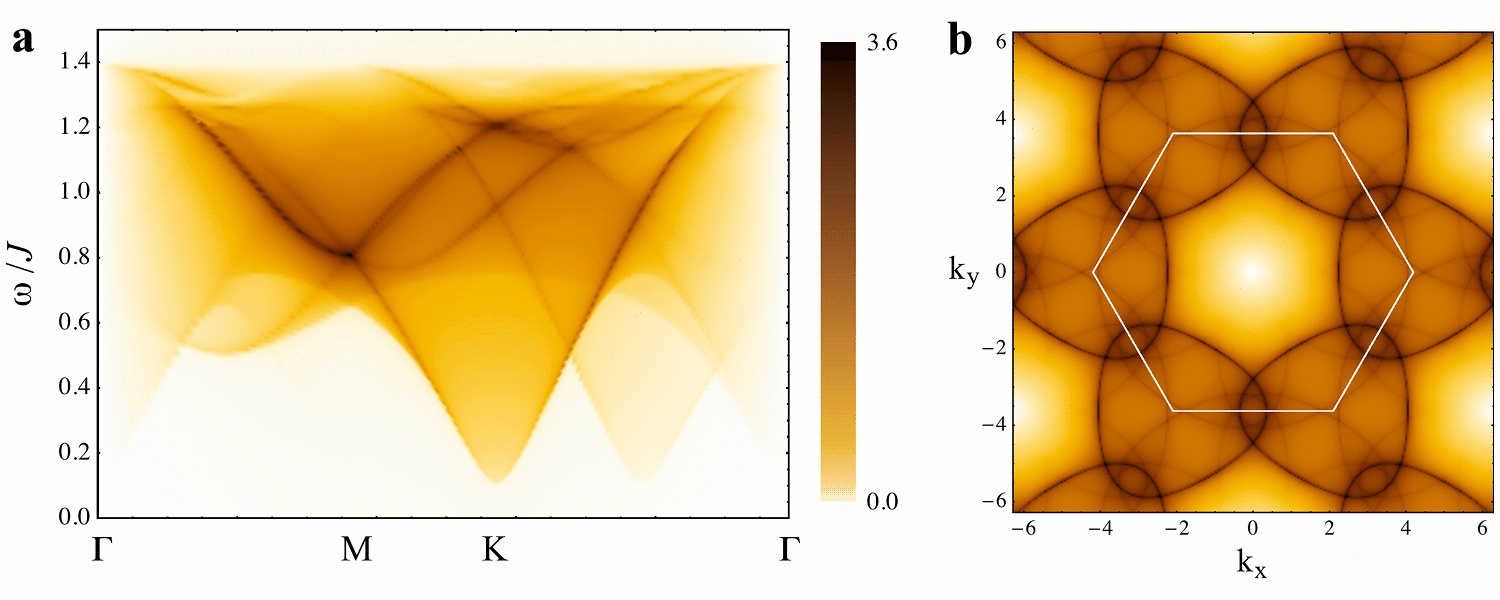}
\end{center}
\caption{Density plots of the dynamical structure factor $S(\k,\omega)$ for the $Q_1=-Q_2$ state with non-interacting spinons. Panel (a) shows $S(\k,\omega)$ as function of frequency and momenta along three high symmetry directions between the $\Gamma, M$ and $K$ points of the extended Brillouin zone. In panel (b) the structure factor is plotted  as function of momenta at fixed frequency $\omega/J=0.9$. The extended Brillouin zone is indicated by the white hexagon.}
\label{figQ1mQ2}
\end{figure}

\subsection{Discussion of the spinon-vison interaction}

In this section we discuss our choice of the Hamiltonian \eqref{H_SV} in more detail. Although we are confident that our simplified model in Equ.~\eqref{H_SV} captures the essential physics, it assumes that the hopping amplitudes of the visons are identically zero. In reality this is not the case, however, and the localization of visons arises due to the destructive interference of different hopping paths. Ideally, the bare vison action would thus take the form
\begin{equation}
S_v = \sum_{i,j, \Omega} \phi_{i,\Omega} \big[ (\Omega^2+m^2) \delta_{i,j} + J_{ij} \big] \phi_{j,-\Omega}
\label{S_V}
\end{equation}
where $m$ is a mass term and the vison hopping amplitudes $|J_{ij}| = J$ are subject to the full-frustration condition $\prod_\text{plaq.} \text{sign}(J_{ij}) = -1$ \cite{Huh}. In order to fulfill this constraint it is necessary to choose a gauge for the signs of the $J_{ij}$'s, which in turn requires to work with an enlarged unit cell. The smallest unit cell where the gauge can be fixed has twice the size of the elementary unit cell, i.e.~6 instead of 3 sites. This in turn would require us to work with spinon propagators that are $12 \times 12$ instead of $6\times6$ matrices, which would turn our calculations into a very cumbersome endeavor. For this reason we restricted our efforts to the simplified model in Equ.~\eqref{H_SV}, which can be implemented on the basic three-site unit cell.

Nevertheless, we would like to estimate the accuracy of our simplified model. For this reason we replace the bare spinon part in Equ.~\eqref{H_SV} with \eqref{S_V}, project to the lowest vison band, integrate out the visons and compare the resulting spinon-spinon interaction with the corresponding result for our original model. We performed this computation on a finite, 12-site lattice with periodic boundary conditions.

We study the effect of a gauge-invariant local energy-energy coupling between {\it only} the visons at the center of the {\it triangular} kagome plaquettes and spinons on the surrounding bonds. This can be schematically written as follows,
\beqn
S_{sv}&=&g_0\sum_{i}\Delta_v \phi_i^2\sum_{lm\in\bigtriangleup} Q_{lm}B_{lm} M_{ilm} ~\tn{where},\\
B_{lm}&=&\varepsilon_{\alpha\beta}b_{l\alpha}b_{m\beta}+\tn{h.c.}, \label{Blm}
\eeqn
represent the bond variables (arranged such that for any given bond $l<m$), $M_{ilm}$ is a matrix element that restricts only the triangular bonds on the kagome lattice to couple to the visons at their center (we have already assumed that $Q_{lm}$ take real values) and $\Delta_v$ is the vison gap.

Let us now introduce an operator, $\mathbb{P}$, which projects to the lowest vison band with energy $\Delta_v(=\sqrt{m^2-\sqrt{6}})$ such that $\mathbb{P}^2=\mathbb{P}$. Upon integrating out the visons we get,
\beqn
S_{\tn{eff}}&=&-\frac{1}{2}\tn{Tr}\log[\mathbb{P}(\mathbb{S}^{-1}\mathbb{J}\mathbb{S}+g_0\Delta_v\mathbb{S}^{-1}\mathbb{V}\mathbb{S})\mathbb{P}]=-\frac{1}{2}\tn{Tr}\log[\mathbb{P}\mathbb{S}^{-1}\mathbb{J}\mathbb{S}\mathbb{P}(1+g_0\Delta_v(\mathbb{P}\mathbb{S}^{-1}\mathbb{J}\mathbb{S}\mathbb{P})^{-1}\mathbb{P}\mathbb{S}^{-1}\mathbb{V}\mathbb{S}\mathbb{P})],
\eeqn
where $\mathbb{J}$, $\mathbb{V}$ are the matrices corresponding to $S_v$, $S_{sv}$ and $\mathbb{S}$ is the similarity transformation that diagonalizes $\mathbb{J}$. If we now expand in small $g_0$ and retain the lowest order non-trivial term, one obtains the vison mediated interaction to be,
\beq
S_{\tn{eff}}=\frac{g_0^2\Delta_v^2}{4}\sum_{\omega_1}\tn{Tr}[\mathbb{G}_{\omega_1}.\mathbb{G}_{\omega-\omega_1}],\tn{where}~\mathbb{G}=(\mathbb{P}\mathbb{S}^{-1}\mathbb{J}\mathbb{S}\mathbb{P})^{-1}\mathbb{P}\mathbb{S}^{-1}\mathbb{V}\mathbb{S}\mathbb{P}.
\eeq

\begin{figure}
\begin{center}
\includegraphics[width=0.4 \columnwidth]{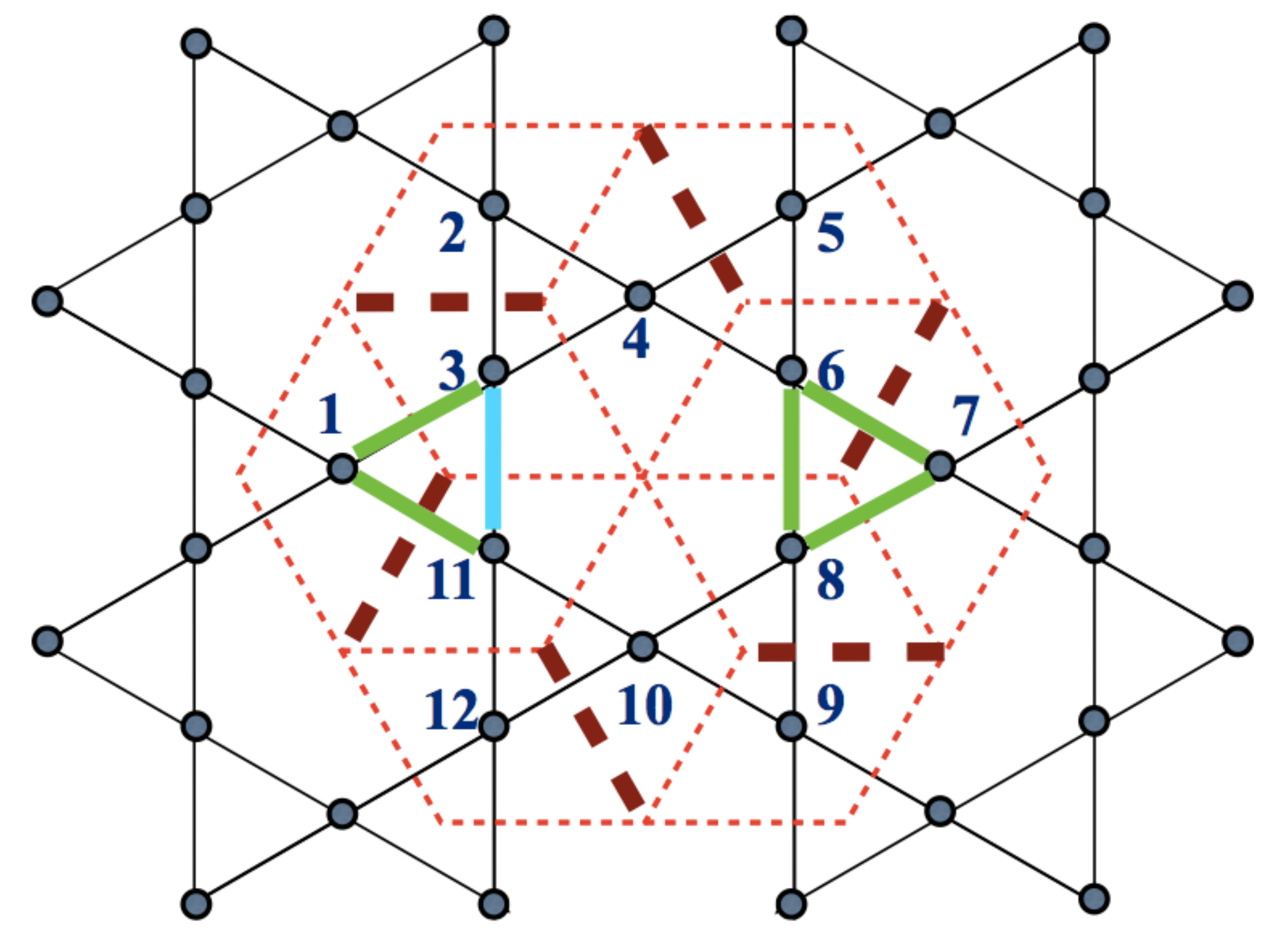}
\end{center}
\caption{Vison induced interaction between spinons. Red dotted lines indicate the 12-site Dice lattice with periodic boundary conditions. We fix the gauge by setting $J_{ij}=-1$ on the thick red dashed bonds. The thick green and blue bonds on the kagome lattice represent bond operators defined in Equ.~\eqref{Blm}. After integrating out the visons, the blue bond interacts with all green bonds. }
\label{figindint}
\end{figure}

If we focus on one particular bond-operator on the kagome lattice and ask which other bonds it couples to (for the $Q_1=Q_2=1$ ansatz), we get for instance for the bond $B_{311}$ (shown as thick blue line in fig.\ref{figindint}),
\beq
S_{\tn{eff}}|_{B_{311}}=\frac{g_0^2\Delta_v}{32(\omega^2+4\Delta_v^2)}B_{311}\bigg[B_{13}+B_{67}-B_{68}+B_{78}-B_{111}\bigg].
\eeq
This is shown schematically in fig.\ref{figindint}. The blue bond interacts with all the green bonds after integrating out the visons. In our simplified model \eqref{H_SV} the blue bond only interacts with the bonds on the same triangle. Accordingly, the simplified Hamiltonian represents a short-range truncation of the induced spinon-spinon interaction.

\subsection{Vertex corrections}

Here we calculate the lowest order vertex correction to the one-loop expression for the spin susceptibility and show that it is small at the $M$ and $K$ points for the $Q_1=Q_2$ state. The diagram corresponding to this vertex correction is shown in Fig.~\ref{figdiags}. Using the shorthand notation $q = (i\Omega_q,\q)$ it takes the explicit form
\begin{eqnarray}
\chi^>(\k,i \omega) &=& \frac{1}{2} \sum_{q,p} G^{}_\ell(p) \, G^{}_m(k-p) \, G_n(q) \, G_o(k-q) \,  U^*_{i\ell}(\p) V^*_{im}(\k-\p) V_{jn}(\q) U_{jo}(\k-\q) \notag \\
&& \times \Bigg[ \frac{3 \, \lambda^u_{nl}(\q,\p) \lambda^{u*}_{mo}(\k-\p,\k-\q)}{i \Omega_p-i \Omega_q+2 \Delta_v}  + \frac{\lambda^v_{nl}(\q,\p) \lambda^{v*}_{mo}(\k-\p,\k-\q)}{i \Omega_p-i \Omega_q+2 \Delta_v}  \Bigg] \ .
 \end{eqnarray}
The $3\times3$ matrices $U$ and $V$, which form the Bogoliubov rotation matrix that diagonalizes the mean-field spinon Hamiltonian, have been introduced in the main text. Sums over the sublattice indices $i,j,l,m,n,o$ are implicit. $\lambda^u$ and $\lambda^v$ denote the diagonal and off-diagonal $3 \times3$ blocks of the $6\times6$ spinon-vison interaction vertex $\lambda$. Again we only show the terms which give rise to a contribution at positive external frequencies after analytic continuation.
Results of a numerical evaluation of this correction at a spinon-vison coupling $g_0=0.6$ are shown in Fig.~\ref{figvc1}, where vertex corrections give a contribution to the structure factor on the order of a few percent and tend to flatten it at intermediate frequencies.
Note that at small external frequencies higher order vertex corrections play no role. This is because they correspond processes where the spinons excite multiple gapped visons and thus give a contribution to $S(\k,\omega)$ only at frequencies $\omega > 2 \Delta_s + n \Delta_v$, where $n$ is the number visons in the intermediate state.

\begin{figure}
\begin{center}
\includegraphics[width=0.4 \columnwidth]{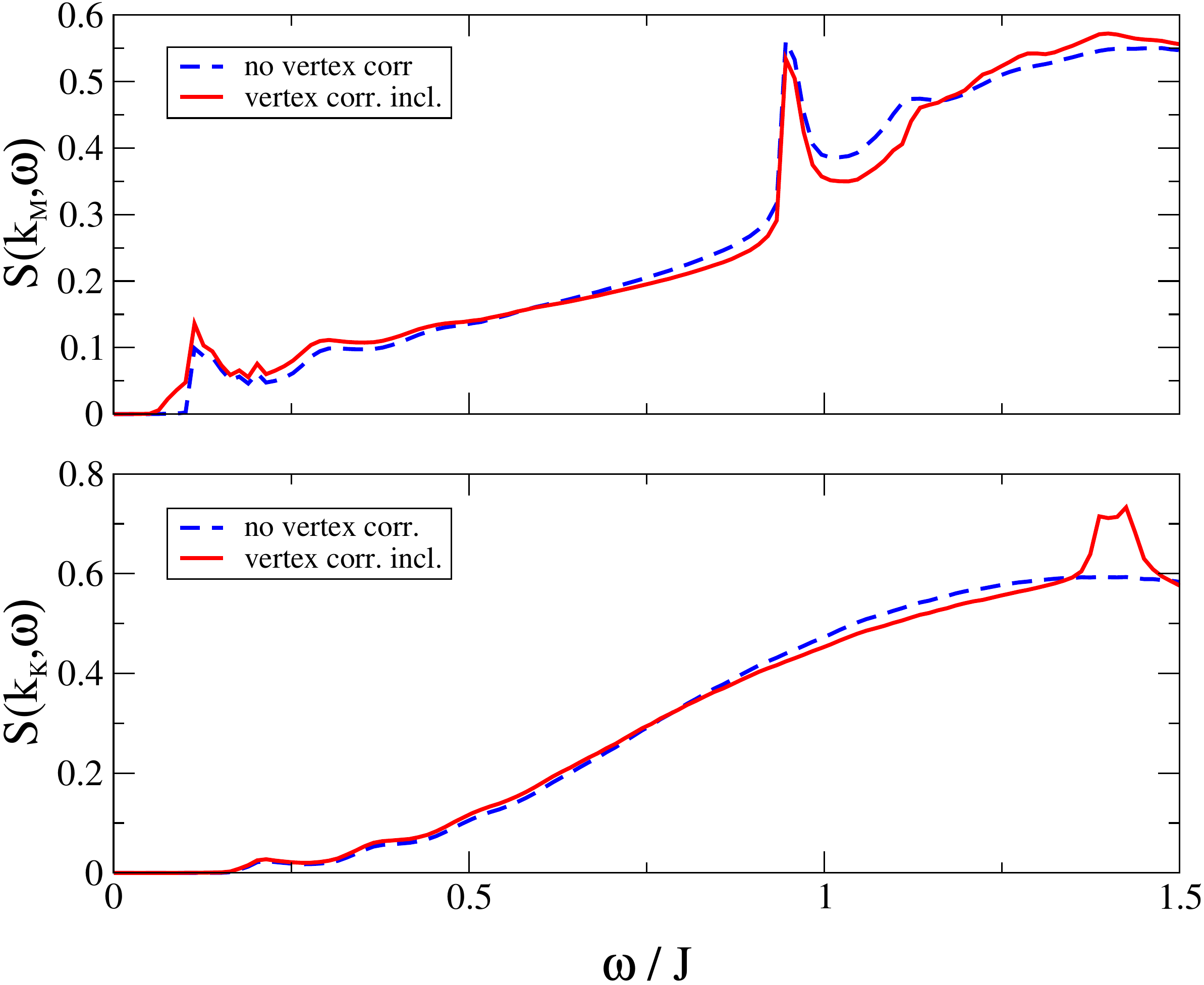}
\end{center}
\caption{$S(\k,\omega)$ at the M (top) and K (bottom) points for the $Q_1=Q_2$ state with and without vertex corrections. Interaction strength $g_0=0.6$, all other parameters as in 
Fig.~\ref{fig1}.}
\label{figvc1}
\end{figure}

\subsection{Spinon spectral function}
\label{secSpecfunc}

An analysis of the spinon spectral function $A(\k,\omega) = - \sum_\ell \text{Im} G_\ell(\k,\omega)$ allows us to evaluate several approximation schemes for calculating the spinon self-energy. In Fig.~\ref{figspecfunc} we compare the spectral function of non-interacting spinons with its interacting counterparts. We plot the spectral function of interacting spinons at three different approximation levels for the spinon self-energy $\Sigma(\k,\omega)$: non-selfconsistent, non-selfconsistent but neglecting the real part, and a selfconsistent calculation of $\Sigma(\k,\omega)$.

The non-selfconsistent computation shown in Fig.~\ref{figspecfunc}(b) gives rise to unphysical sharp dispersing excitations above the highest bare spinon band. These sharp excitations are an artifact of the non-selfconsistent approximation, because they would immediately decay via vison pair-production.
It is easy to see that a large real part of the spinon self-energy is responsible for these features. Consequently, this problem can be fixed by neglecting the real part of the self-energy completely, arguing that renormalizations of the spinon dispersion can be absorbed into the parameters $Q_1$ and the Lagrange multiplier $\lambda$. One has to keep in mind, however, that this approximation violates sum-rules (in particular, the integrated spinon spectral weight is no longer unity). The resulting spectral function is shown in Fig.~\ref{figspecfunc}(c). Note that flat spinon band is still sharp at this approximation level. This is because the band stays flat and the band-gap to the lower spinon bands is larger than twice the vison gap, thus neither intraband nor interband vison-pair production processes are possible.

Finally, results of a self-consistent calculation are shown in Fig.~\ref{figspecfunc}(d). The only qualitative difference to the previously discussed approximation is a broadened flat spinon band away from the $\Gamma$ point. This is mainly because interband decay processes are now allowed, since incoherent spinon excitations are possible inside the bandgap.

\begin{figure}
\begin{center}
\includegraphics[width=\columnwidth]{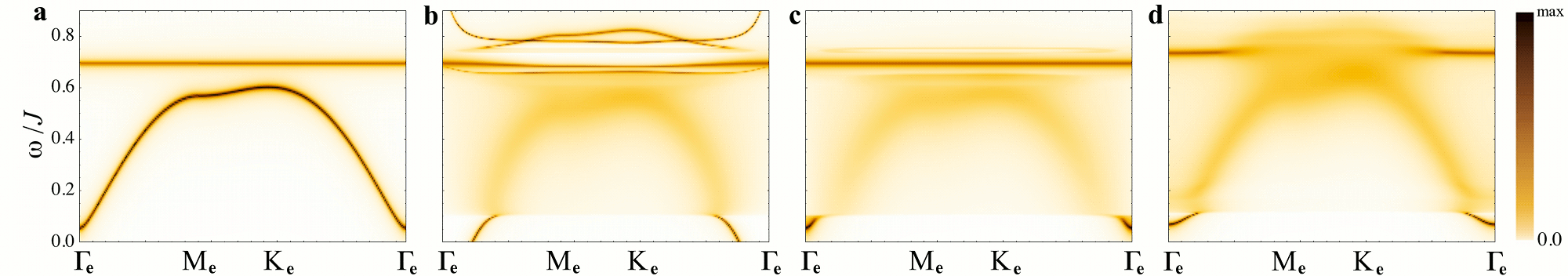}
\end{center}
\caption{Density plots of the spinon spectral function at different levels of approximation for the $Q_1=Q_2$ spin liquid state, plotted along the high symmetry directions between the $\Gamma_e, M_e$ and $K_e$ points of the elementary Brillouin zone (sharp delta-function peaks are broadened for better visibility). (a) non-interacting spinons; (b) spinon-vison interaction $g_0=0.2$, self-energy calculated non-selfconsistently; (c) spinon-vison interaction $g_0=0.2$, self-energy calculated non-selfconsistently, real part of self-energy neglected; (d) spinon-vison interaction $g_0=0.2$, self-consistent spinon self-energy. Note that the lower spinon band is doubly degenerate. See text for a discussion.}
\label{figspecfunc}
\end{figure}

\subsection{Dynamic structure factor in the non-selfconsistent approximation}

In this section we briefly report calculations of the dynamic structure factor using the non-selfonsistently computed spinon self-energy without the real part, as discussed in the preceding section. The results are shown in Fig.~\ref{figSF2} for two different spinon-vison couplings. Compared to the self-consistent results discussed in the main text, two qualitative differences are noteworthy. First, the presence of the sharp flat spinon band in the non-selfconsistent approximation gives rise to a prominent peak in the dynamic structure factor at an energy around $\omega \simeq 0.75 J$. Secondly, the onset of the dynamic structure factor at low energies close to the $M$ point is dominated by a rather sharp peak as well. In comparison, the self-consistent calculation exhibits a smaller peak at the onset of the spectrum around the $M$ point, but it is distributed over a wider range of momenta. We attribute this to a sizable renormalization of the spinon dispersion due to the interaction with visons, which leads to a flattening around the minimum of the lowest spinon band, as seen in Fig.~\ref{figspecfunc}(d). The non-selfconsistent approximation does not account for this flattening.

\begin{figure}
\begin{center}
\includegraphics[width=0.8\columnwidth]{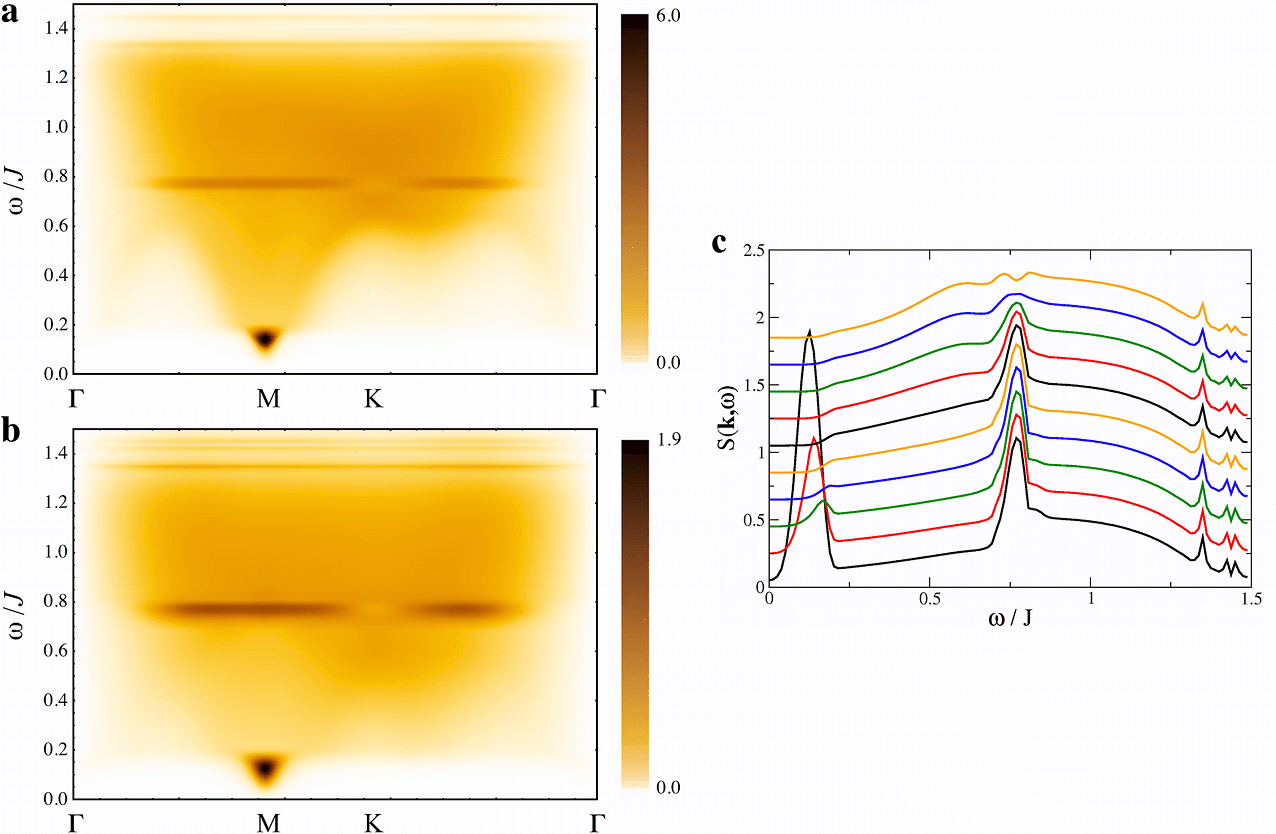}
\end{center}
\caption{One loop result for the dynamic structure factor, computed using the non-selfconsistent spinon self-energy without the real part.  (a): spinon-vison interaction $g_0=0.2$, (b): $g_0=0.4$, (c): details of the structure factor at $g_0=0.4$ between the $M$ (bottom curve) and the $K$ point (top curve). Other parameters as in Fig.~\ref{fig1}. }
\label{figSF2}
\end{figure}

\end{widetext}

\end{document}